\documentclass[journal,twocolumn,10pt]{IEEEtran}
\usepackage{textcomp}
\usepackage{stfloats}
\usepackage{url}
\usepackage{graphicx}
\usepackage{verbatim}
\usepackage{upgreek}
\usepackage{amssymb,amsmath}
\usepackage{color}
\usepackage{epstopdf}
\usepackage{bm}
\usepackage{cite}
\usepackage{array,color}
\usepackage{algorithm}
\usepackage{algpseudocode}
\usepackage{amsmath}
\usepackage{amsfonts}
\usepackage{amsthm}
\usepackage{graphics}
\usepackage{epsfig}
\usepackage{soul}
\usepackage{gensymb}
\soulregister\cite7
\soulregister\ref7
\usepackage{subfigure}
\usepackage{setspace}

\begin{document}

\title{Air-Ground Integrated Sensing and Communications: Opportunities and Challenges}

\author{Zesong Fei,~\IEEEmembership{Senior~Member,~IEEE}, Xinyi Wang,~\IEEEmembership{Graduate~Student~Member,~IEEE},  \\Nan Wu,~\IEEEmembership{Member,~IEEE}, Jingxuan Huang, and J. Andrew Zhang,~\IEEEmembership{Senior~Member,~IEEE}

\thanks{This work was supported in part by National Key Research and Development Program of China (No.2021YFB2900600), and in part by National Natural Science Foundation of China under Grant U20B2039. (Corresponding author: Nan Wu)}
\thanks{Zesong Fei, Xinyi Wang, Nan Wu, and Jingxuan Huang are with the School of Information and Electronics, Beijing Institute of Technology, Beijing 100081, China (E-mail: feizesong@bit.edu.cn, bit\_wangxy@163.com, wunan@bit.edu.cn, jxhbit@gmail.com).}
\thanks{J. Andrew Zhang is with the School of Electrical and Data Engineering, University of Technology Sydney, NSW, Australia 2007 (E-mail: Andrew.Zhang@uts.edu.au).}
}



\maketitle

\begin{abstract}
The air-ground integrated sensing and communications (AG-ISAC) network, which consists of unmanned aerial vehicles (UAVs) and ground terrestrial networks, offers unique capabilities and demands special design techniques. In this article, we   provide a review on AG-ISAC, by introducing UAVs as ``relay'' nodes for both communications and sensing to resolve the power and computation constraints on UAVs. We first introduce an AG-ISAC framework, including the system architecture and protocol. Four potential use cases are then discussed, with the analysis on the characteristics and merits of AG-ISAC networks. The research on several critical techniques for AG-ISAC is then discussed. Finally, we present our vision of the challenges and future research directions for AG-ISAC, to facilitate the advancement of the technology.
\end{abstract}


\section{Introduction}

\IEEEPARstart{W}{ith} the vision of ``Intelligent Internet of Everything'', next-generation wireless networks, e.g., beyond 5G (B5G) and 6G, have been envisioned as key enablers for various applications \cite{Saad2020MNET}. One of the major characteristics of the future network is the dynamic topology and various complex environments. As a result, unmanned aerial vehicles (UAVs) are expected to play key roles in a wide range of scenarios in future networks due to their advantages of high mobility and fast deployment, particularly when they function as a network. In view of this, there have been numerous works on improving the communication \cite{Li2019IOT} and sensing \cite{Xiang2019MGRS} performance of UAV networks. Nevertheless, the size, weight, and power (SWAP) constraints make it challenging to install both communication and sensing systems on UAVs. Moreover, deploying a large number of UAVs, in which some provide communication services while the others perform sensing, will not only introduce co-channel interference, but also increase the resource consumption \cite{Wang2021TCOM}.

The integrated sensing and communications (ISAC) technique potentially provides a great solution to effectively enabling both functions in a UAV network. The ISAC technique is able to improve the spectral efficiency, hardware efficiency, as well as information processing efficiency, and therefore, has been viewed as a major candidate for 6G networks \cite{liu2021integrated}. Applying ISAC to UAVs, we can not only minimize payload and resource usage, but also improve the sensing performance through cooperation between communications and sensing.

Nevertheless, considering power consumption and cost of UAVs, it may be inefficient to implement computation-intensive processing on UAVs, e.g., high-resolution localization or trajectory design, which requires huge amount of computing resources. To overcome such difficulties, mobile edge computing (MEC) \cite{Mao2017survey} can be applied to offload the computation to, e.g., ground mobile base-stations (BSs). By transferring the computation burden to the edge network with low access load and transmission delay, the MEC technique is able to relieve the computation burden of UAVs.

Through building up an air-ground integrated network, the terrestrial network can provide edge computing capabilities for UAVs, while the UAVs with ISAC capabilities are able to expand the coverage of the terrestrial network and improve the sensing and communication performance by, e.g., creating Line-of-Sight (LoS) links. Such an air-ground ISAC (AG-ISAC) network is expected to play an important role in both civilian and military applications. Nevertheless, the AG-ISAC network has some distinctive features compared to conventional ground-based ISAC networks and requires special techniques to solve some critical challenges. There have been some solid progress being made to advance the AG-ISAC technique, although many challenges are yet to be solved. In this article, we aim to provide a review of AG-ISAC and shed lights on future development of the technology. We first present its system architecture for AG-ISAC network, together with the frame structure, including system setup, processing chains, an exemplified protocol, and typical applications. We then elaborate several critical techniques, illustrating and comparing solutions. Finally, we highlight the technical challenges and future research directions.

\section{System Architecture}

\begin{figure*}
 \centering
 \includegraphics[width=7 in]{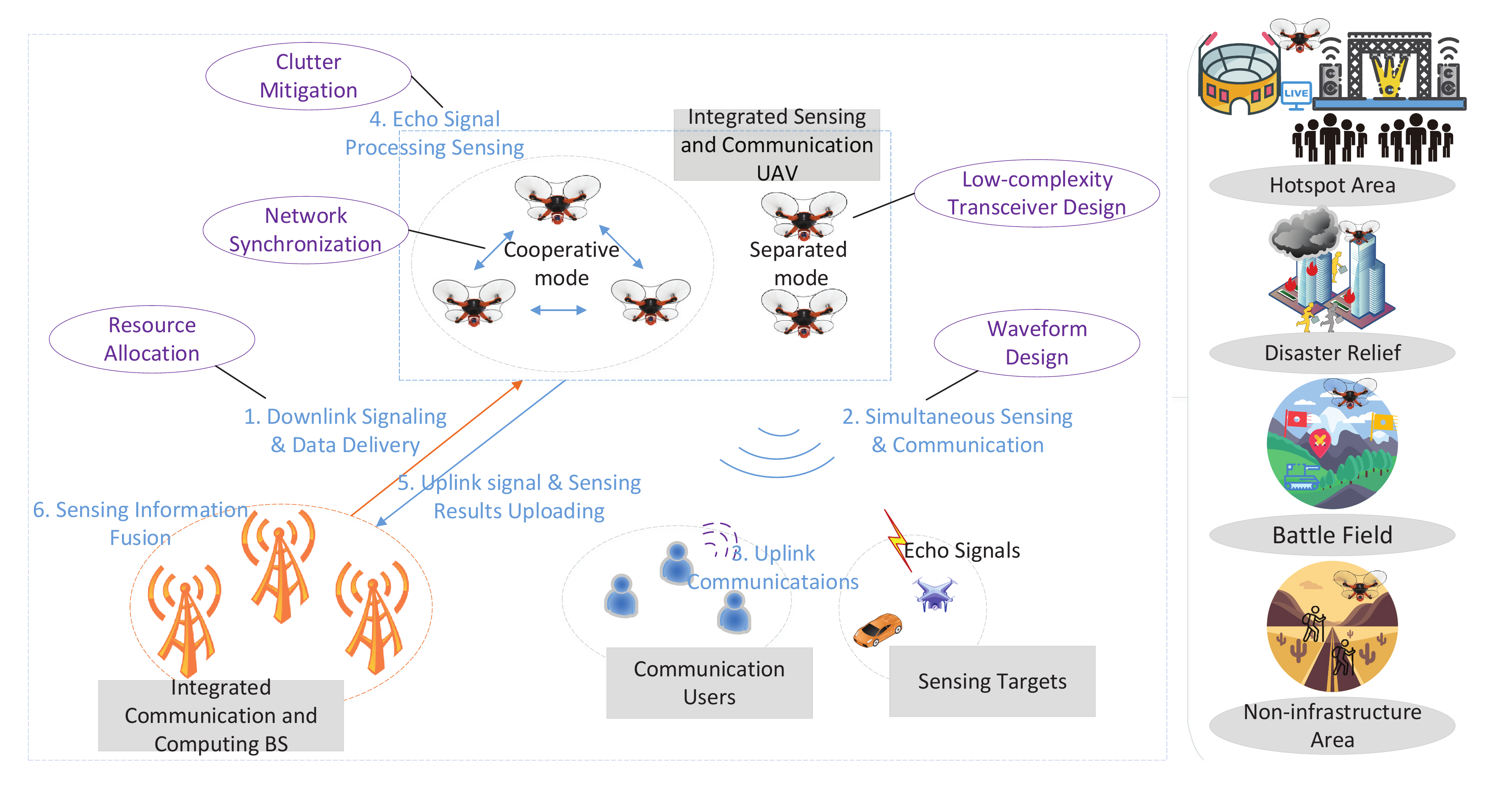}
 \caption{Architecture and typical applications for AG-ISAC Enabled Future Network. (The numbered text represents the main processing stages in the network. Text in the ellipse denotes key technologies associated with these stages. They will be elaborated in following sections.)}
 \label{system_model}
\end{figure*}

Fig. \ref{system_model} describes the system architecture of the AG-ISAC network, where multiple UAVs, connected with the terrestrial BSs via wireless backhaul, are deployed to provide flexible and seamless coverage for communication users. Meanwhile, the UAVs can also execute sensing tasks, such as target sensing or tracking based on the echo signals. To handle the heavy computation tasks in high-accuracy sensing, UAVs can offload the locally processed sensing results to BSs, which then perform sensing information fusion. In this sense, the BS with high computing capability and UAVs with high mobility can be respectively viewed as ``brain'' and ``sense organs'' in human bodies, while the communication signals between BSs and UAVs serve as ``nerve impulses''. In addition to the sensing task offloading, UAVs can also help offload heavy tasks from ground users to BSs via uplink communications. Thanks to the fully controllable mobility of UAVs, the AG-ISAC network is expected to play important roles in future civilian and military applications. Four potential scenarios are illustrated in Fig. \ref{system_model} and elaborated in Table \ref{use-cases}. Note that while providing communication services in non-infrastructure area, UAVs can either constitute a multi-hop relay network to relay communication signals from a remote BS, or serve as BSs and use, e.g., Satellite communications as backhaul connections, via directional antennas.

\begin{table*}
\centering
\caption{Potential Use Cases of AG-ISAC Network}
\setstretch{1.2}
\label{use-cases}
\resizebox{\textwidth}{!}
{\begin{tabular}{|l|l|l|}
\hline
  \multicolumn{1}{|c|}{\textbf{Use Cases}} &
  \multicolumn{1}{c|}{\textbf{Characteristics}} &
  \multicolumn{1}{c|}{\textbf{Merits of Deploying AG-ISAC Network}} \\ \hline
  Hotspot Area &
  \begin{tabular}[c]{@{}l@{}}$\bullet$ Highly dynamic and heterogeneous \\ $\bullet$ Composed of both static and mobile nodes  \end{tabular} &
  \begin{tabular}[c]{@{}l@{}}$\bullet$ Constantly sense the dynamic environment and topology \\ $\bullet$ Swiftly react to the varied demands \\ $\bullet$  Provide high-throughput communications \end{tabular} \\ \hline
  Disaster Relief &
  \begin{tabular}[c]{@{}l@{}} $\bullet$ Terrestrial network may be destroyed \\ $\bullet$ Dangerous for search \& rescue operation \end{tabular} & \begin{tabular}[c]{@{}l@{}} $\bullet$ Provide robust and fast emergency communications \\ $\bullet$ Improve rescue efficiency through precise localization \\ $\bullet$ Provide damage assessment via sensing data analysis \end{tabular}
   \\ \hline
  Battle Field &
  \begin{tabular}[c]{@{}l@{}} $\bullet$ Require constant invasion monitoring \\ $\bullet$ Plenty of latency-critical computing tasks \end{tabular} &
  \begin{tabular}[c]{@{}l@{}} $\bullet$ Simultaneous target sensing and information sharing \\ $\bullet$ Improve task execution efficiency via cooperation \\ $\bullet$ Provide low-latency task offloading \end{tabular}  \\ \hline
  Non-Infrastructure Area &
  \begin{tabular}[c]{@{}l@{}} $\bullet$ Require continuous environment monitoring \\ $\bullet$ Unworthy and impractical to deploy infrastructures \end{tabular} & 
  \begin{tabular}[c]{@{}l@{}} $\bullet$ Provide low-cast communication and sensing services \\ $\bullet$ Provide high-throughput measurement data uploading \end{tabular} \\ \hline
\end{tabular}
}
\end{table*}

\begin{figure*}
 \centering
 \includegraphics[width=7 in]{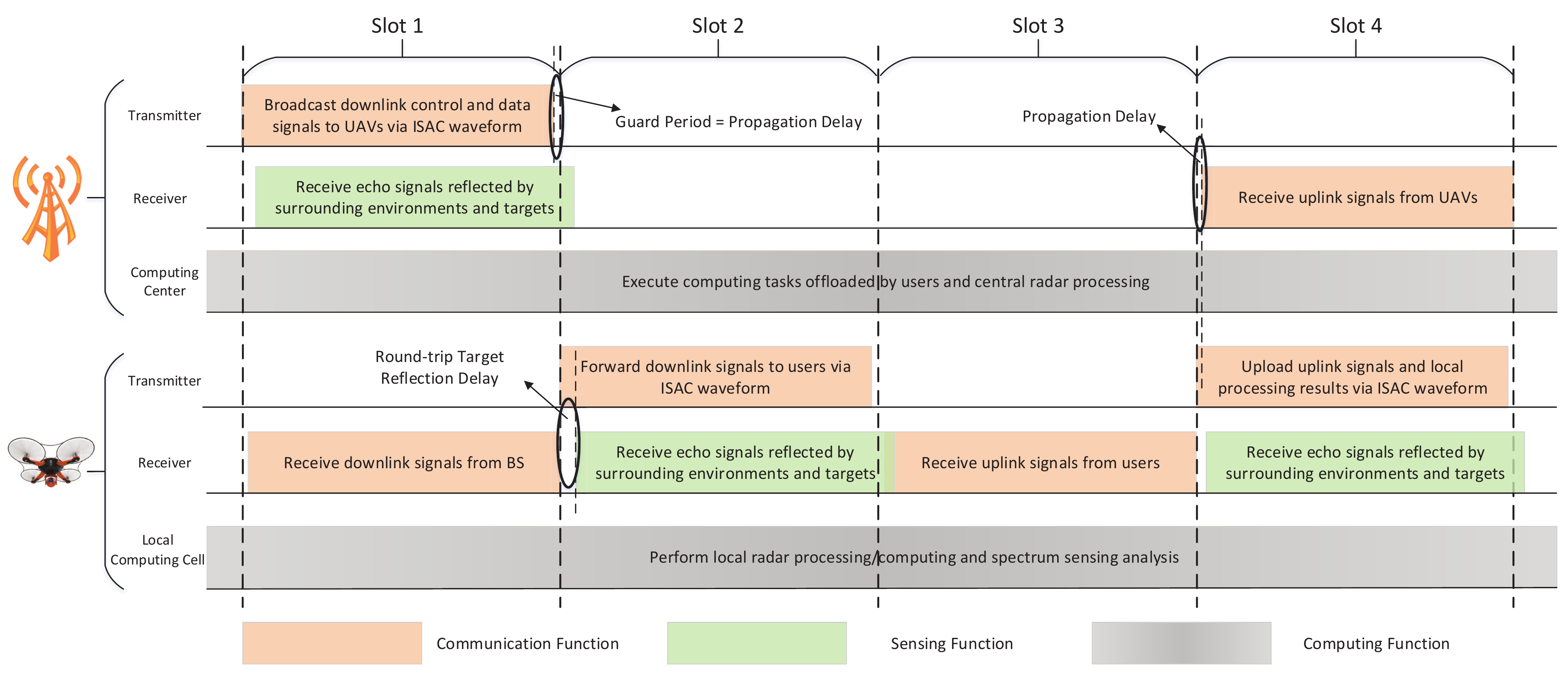}
 \caption{Protocol of the AG-ISAC network.}
 \label{protocol}
\end{figure*}

In the AG-ISAC network, the roles of UAVs are twofold. On one hand, the UAVs act as relays to provide robust and enhanced communication services for ground users. On the other hand, UAVs extend the vision of BSs, thereby enlarging the sensing range. To realize these two functions, a set of stages are combined to form a complete processing, as illustrated in Fig. \ref{system_model}. These stages are supported by carefully designed protocols. One of such a protocol is illustrated in Fig. \ref{protocol}, where the communication function follows the principle of ``BS-UAV-user'' two-hop transmission, and the BS performs sensing information fusion based on UAV's pre-processed local sensing results. Specifically, each frame is divided into four slots. In the first slot, the BS broadcasts downlink control and data signals to UAVs via ISAC waveform so that the environmental information can be obtained based on echo signals. Then, in the second slot, the UAVs forward the downlink communication signals to users via ISAC signals, and simultaneously receive echo signals reflected by targets. Due to the signal propagation delay, a guarding period is reserved at slot 1 to prevent the reception process of UAVs from being affected by its transmission process. Upon receiving echo signals, the UAVs perform local preprocessing, e.g., removing unwanted clutter and matched filtering (also known as pulse compression in radar sensing), so that the size of data to be offloaded to BS is reduced. Subsequently, the UAVs receive uplink signals from users and upload uplink communication signals as well as local processing results to BS. Based on the sensing information from UAVs, the BS perform central processing/computing to achieve high-quality sensing, e.g., constant false alarm rate (CFAR) detection or target recognition. Note that before sensing information is uploaded, pre-processing is performed by UAVs such that the information to be uploaded can be compressed; thus, with the aid of artificial intelligence enabled information fusion technique, the sensing task offloading will not incur a large processing delay.

While executing sensing tasks, the UAVs can work in two modes, i.e., separated mode and cooperative mode. In the separated mode, each UAV senses the target or environment based on the ISAC waveform transmitted by itself, while the signals transmitted by other UAVs are treated as interference. Although the separated mode is easy to implement, the quality of the received echo signals is limited, and the BS can only perform information-level fusion of sensing results; thus, the sensing performance may not be satisfied. On the other hand, in the cooperative mode, since the signals transmitted by the BS are broadcast to all UAVs, the ISAC signals transmitted by each UAV are known to the others; thus, the UAV network can act as a distributed MIMO system and perform cooperative sensing. Compared with the separated mode, the cooperative mode is able to achieve signal-level sensing information fusion at BSs, thereby improving the sensing performance.

\section{Critical Techniques in Integrated Design}

In this section, we discuss some critical techniques in the AG-ISAC network, as presented in Fig. \ref{system_model}.

\subsection{Resource Allocation}

\begin{figure}
 \centering
 \includegraphics[width=3.4 in]{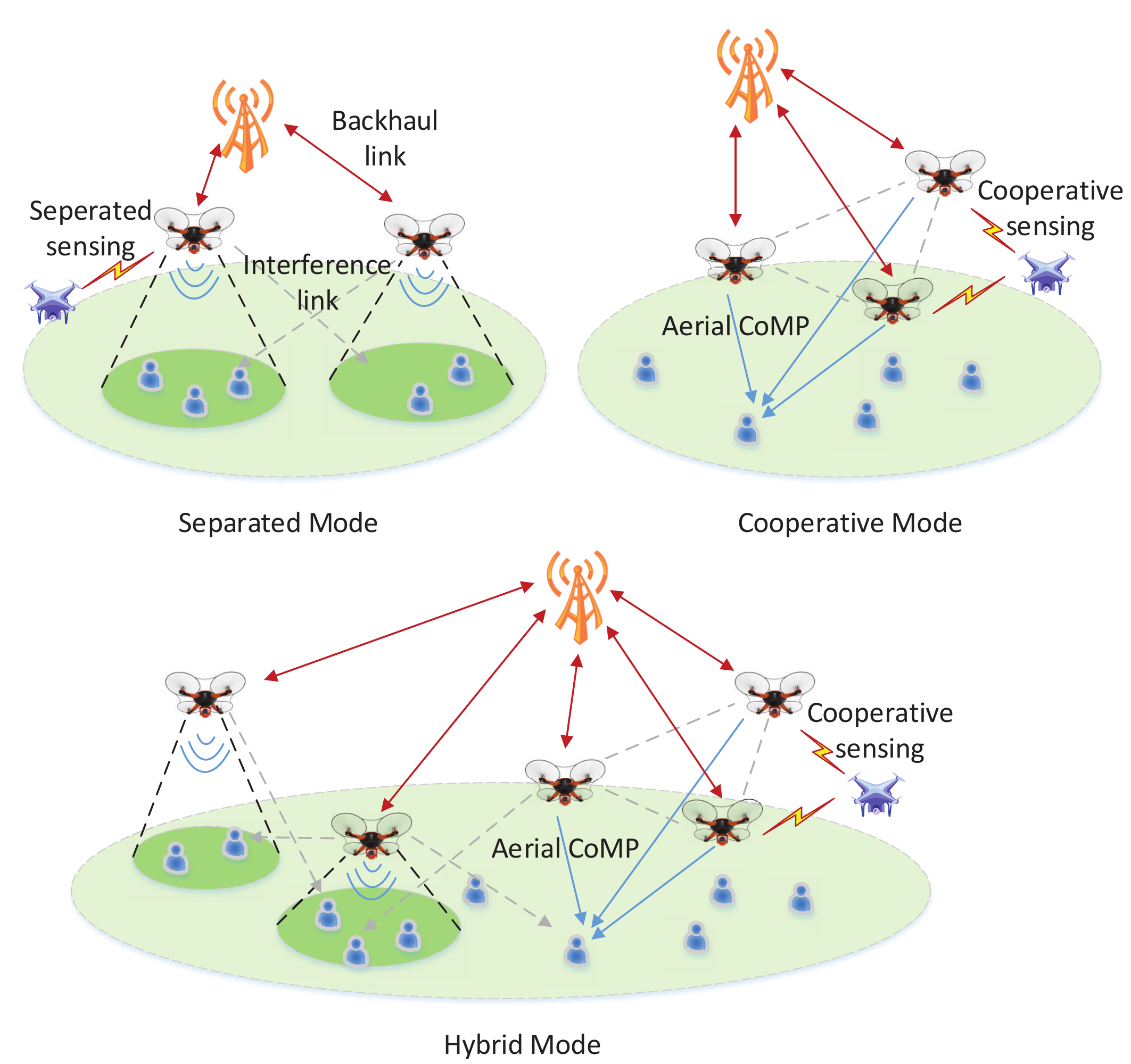}
 \caption{AG-ISAC Network: Separated Mode and Cooperative Mode.}
 \label{mode}
\end{figure}

In the AG-ISAC network, the working modes of UAVs can be divided into separated mode, cooperative mode, and hybrid mode, as shown in Fig. \ref{mode}. Resource allocation in these three modes is quite different, and we discuss them separately next.

\subsubsection{Separated Mode}

While the UAVs execute communication and sensing tasks in a separate manner, the signals transmitted by the other co-channel UAVs will cause severe interference due to the strong line-of-sight (LoS) dominant air-air and air-ground channels. In addition, from the perspective of sensing, the echo signals in radar based sensing experience two-way path loss, while the interference only experiences one-way path loss. Therefore, inappropriate resource allocation will increase the interference, thus limiting both sensing and communication performances. 

One efficient way for reducing inter-UAV and air-ground interference is to perform multi-domain resource allocation, such as spectrum and power allocation. Conventional power allocation problems with continuous variables can be transformed into convex ones via geometric programming or successive convex approximation techniques. However, spectrum allocation problems contain binary variables; therefore, joint spectrum and power  allocation is a mixed integer programming problem, which is challenging to solve. Fortunately, due to the high altitude of UAVs, the air-air and air-ground channels are dominated by LoS components; therefore, the interference level largely depends on the distance between different nodes, and the interference graph can be constructed based on the geometric topology. Exploiting this property, in \cite{Wang2021SCIS}, two graph theory based algorithms were proposed to jointly optimize the spectrum and power allocation schemes.


\subsubsection{Cooperative Mode}

When working in the cooperative ISAC mode, the signals transmitted by other UAVs can be turned into allies. From the perspective of sensing, multiple UAVs can perform distributed radar sensing . Although the limited on-board computing capability cannot support complicated signal processing, by uploading and fusing locally processed sensing results at the central BS, higher precision and larger coverage can be simultaneously achieved. From the perspective of communications, deploying multiple UAVs enables aerial coordinated multi-point (CoMP) \cite{Liu2019TCOM} transmission, thereby improving the spectral efficiency. In particular, the transmit power of each UAV has to be optimized to pursue large distributed MIMO gain for improving the spatial localization accuracy \cite{Wang2021TCOM}, while the user scheduling strategy shall also be carefully designed, especially when the users are successively served by all UAVs.

\subsubsection{Hybrid Mode}

Although the cooperative mode provides better performance in both sensing and communications compared with the separated mode, it may not be practical to deploy a large amount of cooperative UAVs due to the strict synchronization requirement. Therefore, the UAV network may work in a hybrid mode, i.e. some of them work in cooperative mode, while the others work in separated mode. In this case, the UAVs working in the cooperative mode with shared spectral resource can be viewed as a sub-system with distributed antennas, and the interfering channels can be constructed based on the channels related to each antenna.


\subsection{ISAC Transceiver and Waveform Design}

In AG-ISAC networks, while providing communication services to users, the UAVs also execute sensing tasks. Different from the communication functionality, which aims to convey information to a receiver, the sensing functionality aims to extract information from the echos reflected by targets. As a result, the performance of the two functionalities are typically measured by different metrics. Therefore, the ISAC waveform needs to be carefully designed to achieve a balance  between communications and sensing. To this end, a popular option is to ensure high SINR at communication users under the beampattern constraints \cite{Liu2018TSP}. Another mainstream mechanism is to exploit their weighted sum as the objective function, which leads to a Pareto-optimality of the multi-objective optimization.

\begin{table*}
\centering
\caption{Comparison of LFM, OFDM, and OTFS based ISAC Waveform}
\setstretch{1.2}
\label{waveform-comparison}
\resizebox{\textwidth}{!}
{\begin{tabular}{|l|l|l|}
\hline
  \multicolumn{1}{|c|}{\textbf{Techniques}} &
  \multicolumn{1}{c|}{\textbf{Advantages}} &
  \multicolumn{1}{c|}{\textbf{Disadvantages}} \\ \hline
  LFM based waveform &
  \begin{tabular}[c]{@{}l@{}}$\bullet$ Superior ambiguity function and PAPR performance \\ $\bullet$ Widely used in existing radar systems  \end{tabular} &
  \begin{tabular}[c]{@{}l@{}}$\bullet$ Low data rate while embedding information symbols \end{tabular} \\ \hline
  OFDM based waveform &
  \begin{tabular}[c]{@{}l@{}} $\bullet$ High throughput \\ $\bullet$ Mature low-complexity demodulation schemes \\ $\bullet$ Widely used in existing cellular systems \end{tabular} & \begin{tabular}[c]{@{}l@{}} $\bullet$ Higher PAPR than LTM waveform \\ $\bullet$ Performance degradation in high-mobility scenarios \end{tabular}
   \\ \hline
  OTFS based waveform &
  \begin{tabular}[c]{@{}l@{}} $\bullet$ High throughput \\ $\bullet$ DD-domain demodulation, similar to radar parameter estimation \\ $\bullet$ Robust to high Doppler \end{tabular} &
  \begin{tabular}[c]{@{}l@{}} $\bullet$ Higher PAPR than LTM waveform \\ $\bullet$ Higher implementation complexity than OFDM systems \end{tabular}  \\ \hline
\end{tabular}
}
\end{table*}

Here, some issues specific to the AG-ISAC network are highlighted. On one hand, considering the complexity and cost limits for UAV platforms, it is more desirable to implement hardware-efficient transceivers, e.g., hybrid analog-digital beamforming architecture. On the other hand, to address the high mobility issue of UAVs, it's preferred to adopt the waveform with the ability of resisting the impact of high Doppler frequency. The recently burgeoning orthogonal time frequency space (OTFS) modulation \cite{Wei2021MWC}, which modulates information in the delay-Doppler (DD) domain, shows strong Doppler-resilience and potential of full diversity, thus enabling reliable communications in high-mobility scenarios. Moreover, the characteristics of DD domain channels are closely related to the distances, speeds, and scattering intensities of the objects in the propagation environment. Therefore, OTFS is inherently suitable for AG-ISAC transmission. The comparisons of conventional linear frequency modulation (LFM), orthogonal frequency division modulation, and OTFS based waveforms are presented in Table \ref{waveform-comparison}.

\subsection{Trajectory Design}

Trajectory design also plays an important role in AG-ISAC networks, which need to support both static and mobile communication users. With the aid of sensing and computing capabilities, the AG-ISAC network is able to track and predict the movement of users, thus enabling real-time trajectory design. Besides, as the millimeter-wave communications are becoming increasingly popular in future networks, the severe blockage issue has to be addressed. By exploiting the sensing ability, potential blockages can be detected, and the trajectories of UAVs can be adjusted to maintain a reliable link for users. One major issue in trajectory design is that the UAVs have to sense and avoid collision. With the aid of cooperative sensing, the sensing information can be shared among UAVs such that the sensing coverage of the whole network can be enlarged, thereby enabling more robust trajectory design.

Besides enhancing communication services, trajectory design also help improve the efficiency of UAV-enabled task offloading, especially in multi-user scenarios. UAVs need to decide the order of users to be served by jointly considering the distribution and mobility of users, task priorities, and energy consumption, and thereafter, the trajectories of UAVs are designed with both offloading delay and computation delay taken into consideration.

It should be noted that while deploying UAVs as relays to provide communication or computation task offloading services, it is required to ensure stable channel conditions. This incurs communication connectivity requirements. In addition, since the UAVs are typically resource-limited, the trajectory also needs to be carefully designed to save energy for propulsion. To address this issue, the authors in \cite{Zhang2019TCOM} designed the UAV trajectory subject to practical communication connectivity constraints with BSs based on graph theory and convex optimization.

\subsection{Physical Layer Security}

In ISAC systems, the targets to be sensed are usually non-cooperative, while the ISAC waveform containing communication messages is required to focus on the directions of targets such that the SINR of echo signals is high enough. However, this results in information security issue, especially when the target is an eavesdropper. A popular solution to this problem is to embed artificial noise into ISAC waveform \cite{Su2020TWC}. By jointly designing data beamformers and artificial noise, a trade-off between sufficient probing power towards targets and limited data signal power at the targets can be achieved, such that the eavesdropping can be prevented while achieving effective sensing.

\begin{figure*}
 \centering
 \fbox{\includegraphics[width=7 in]{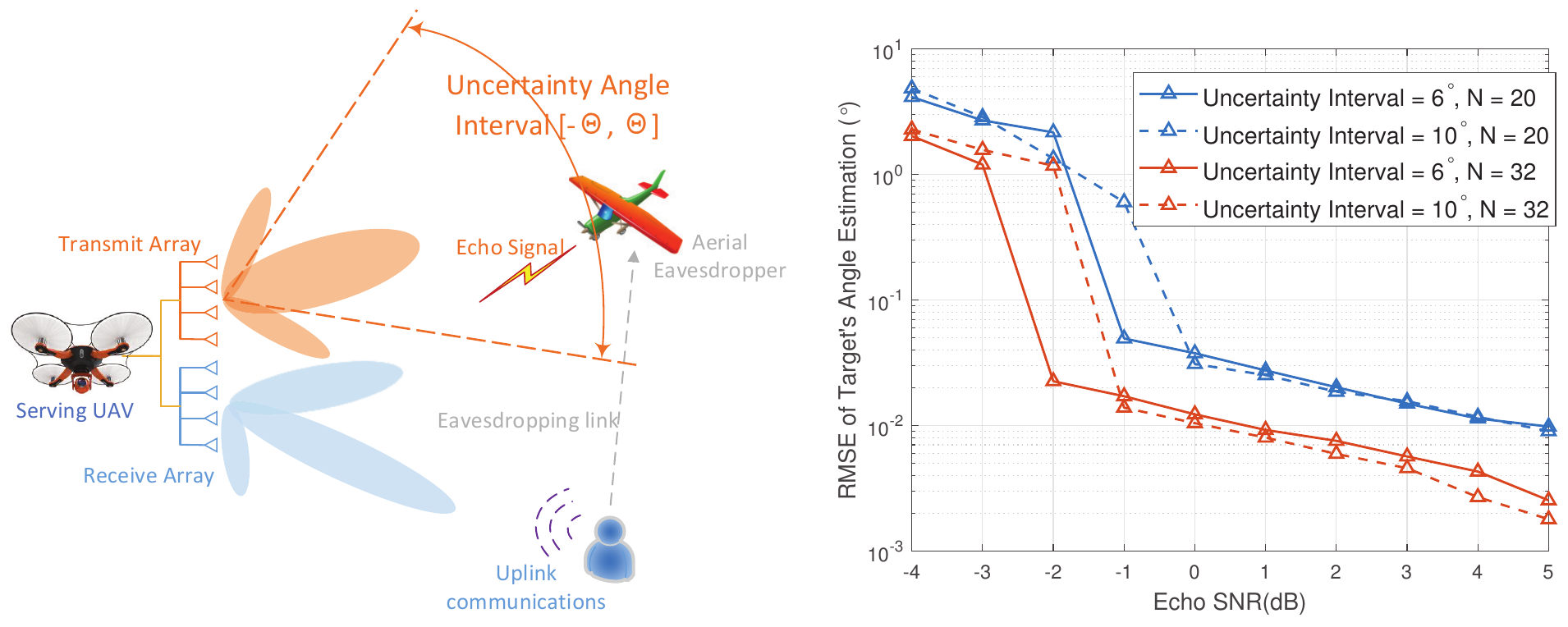}}
 \caption{Sensing-assisted secure uplink communication against the aerial eavesdropper: System model and angle estimation root mean square error (RMSE) performance.}
 \label{secure}
\end{figure*}

Although the ISAC technique faces inherent security issues, it should be noted that the sensing capability also has the potential of enhancing secrecy performance. To be specific, one major challenge in the field of physical layer security is the acquirement of the channel state information (CSI) of eavesdroppers. With the aid of sensing ability, the ISAC transmitters are able to estimate the angles of eavesdroppers, thus constructing the line-of-sight (LoS) component of CSI. We take the sensing-assisted secure uplink communication scenario \cite{Wang2022CL} as an example, as shown in Fig. \ref{secure}. While receiving the uplink signals from the communication user, the serving UAV with sensing capability also transmits sensing signals and estimates the angle information of the aerial eavesdropper. As can be seen, the angle of the aerial eavesdropper can be estimated with high accuracy, thus enabling the construction of partial CSI of the eavesdropper, based on which more efficient jamming can be realized. Moreover, to address the impact of the imperfect CSI introduced by the small-scale fading, robust beamforming design can be utilized by characterizing the imperfect CSI as a Gaussian CSI error model.

Different from the existing studies, which mainly considered the mono-static set-up, such that only the eavesdropper locating in the LoS link can be sensed and jammed, the proposed multi-UAV enabled ISAC network is able to enlarge the coverage of sensing and jamming via cooperation. Therefore, the security level can be further improved.

\section{Challenges and Future Research Directions}

Implementing the proposed AG-ISAC network faces several major challenges, which also mean great research opportunities. We briefly discuss these challenges and potential solutions next.

\subsection{Network Synchronization}

Although the sensing performance provided by one single UAV has been improved thanks to the increased bandwidth at higher frequency bands, e.g., millimeter-wave and tera-herz bands, high-resolution localization requires cooperative sensing, i.e. coherent central processing of the received signals. This requires accurate synchronization among local oscillators of all UAVs performing cooperative sensing. Imperfect synchronization may degrade the ambiguity function of the signals in distributed systems, thus lowering the localization resolution. As specified in \cite{Leyva2021MVT}, centimeter-level resolution accuracy requires the synchronization error among different clocks to be within tens of picoseconds, which is quite a tight synchronization requirement. However, since the UAVs are connected via wireless channels, they cannot be perfectly  synchronized. The resulting phase noise in terms of timing offset and carrier frequency may lead to ambiguity in delay and Doppler estimation. Although some existing synchronization protocols, e.g., precision time protocol (PTP) and master-slave closed-up, can meet a relaxed synchronization requirement, advanced synchronization mechanism are needed in future network.

\subsection{Clutter Mitigation}

In conventional terrestrial wireless communications, BSs may receive plenty of multipath signals reflected by long-period static objects. These signals are exploited to achieve diversity or multiplexing gains, thereby improving the communication performance. However, from the perspective of sensing, the reflections from these static objects are not of interest since they bear little new information. Moreover, such multipath signals may increase the number of sensing parameters to be estimated and degrade the sensing performance. Therefore, such undesirable signals are generally treated as useless clutters. This leads to a multipath exploitation vs. reduction trade-off in terrestrial systems.

Compared with the terrestrial systems, the channels in UAV-enabled air-ground network are dominated by line-of-sight (LoS) components, i.e., the multipath signals contribute less to communication performance. Therefore, it may be more desirable to mitigate the impacts of multipath signals on sensing. There may be two ways for achieving clutter mitigation. On one hand, the ISAC transmit and receive beamformers can be properly designed to maximize the signal-to-clutter-plus-noise ratio (SCNR) of the received echo signals. On the other hand, advanced clutter suppression techniques can be utilized to further improve the SCNR. Although for conventional radar systems, clutter suppression is a widely studied topic, it should be noted that the airborne sensing scenario raises new problems and require radical research on clutter mitigation.

\subsection{Efficient Signal Processing}

The environment of the future network is expected to be highly dynamic, and the channels may be affected by multiple complex factors. As a result, more efficient computing techniques are required to improve the system performance in terms of communication throughput, sensing accuracy, as well as computing delay. Artificial intelligence can be introduced to provide a method for more efficient sensing fusion, such that better vision of targets and environment information can be achieved. For example, the federated learning (FL) framework can be conducted to train intelligent learning models for fusing the sensing information from different UAVs  \cite{Feng2021MNET}. In particular, based on the pre-processing results of the echo signals, each UAV is able to train its local model, i.e., neural network weights. Subsequently, the BS collects the updated model and computes a global model. In this way, the size of information to be shared between UAVs and BS can be significantly reduced, thereby improving the network efficiency. Furthermore, the long short-term memory network can be exploited for future motion state prediction based on the current sensing information.

\subsection{Sensing on Dynamic Platforms}

Although AG-ISAC is able to improve the sensing performance by creating LoS links, the mobility of UAVs also incurs challenges. To be specific, multiple ranging measurements from a single UAV on same targets may lead to large accumulated errors, especially when the UAV is quickly moving. Furthermore, while the sensing results from different UAVs are fused to improve the sensing performance, diverse velocities among different UAVs can lead to different Doppler and accumulated errors, making the fusion more challenging. Therefore, the sensing algorithm has to be designed by considering the mobility of the platforms.

\section{Conclusion}

In this article, we provided a review for AG-ISAC network and technologies, where UAVs are used as the ``relay'' nodes for both communications and sensing in an air-ground network. We introduced an AG-ISAC framework by presenting its system architecture and protocol. Four potential use cases were provided, disclosing special capability requirements for AG-ISAC. We also discussed the research progress on several critical techniques, including resource allocation, waveform design, trajectory design, and physical layer security, which pave the way for advancing the network. Finally, challenges and future research directions were introduced. With its unique capabilities and technical feasibility, AG-ISAC is expected to become one of the key enabling techniques for future communications and sensing applications.

\bibliographystyle{IEEEtran}%
\bibliography{bib/bibfile}

\begin{IEEEbiography}[{\includegraphics[width=1in,height=1.25in,clip,keepaspectratio]{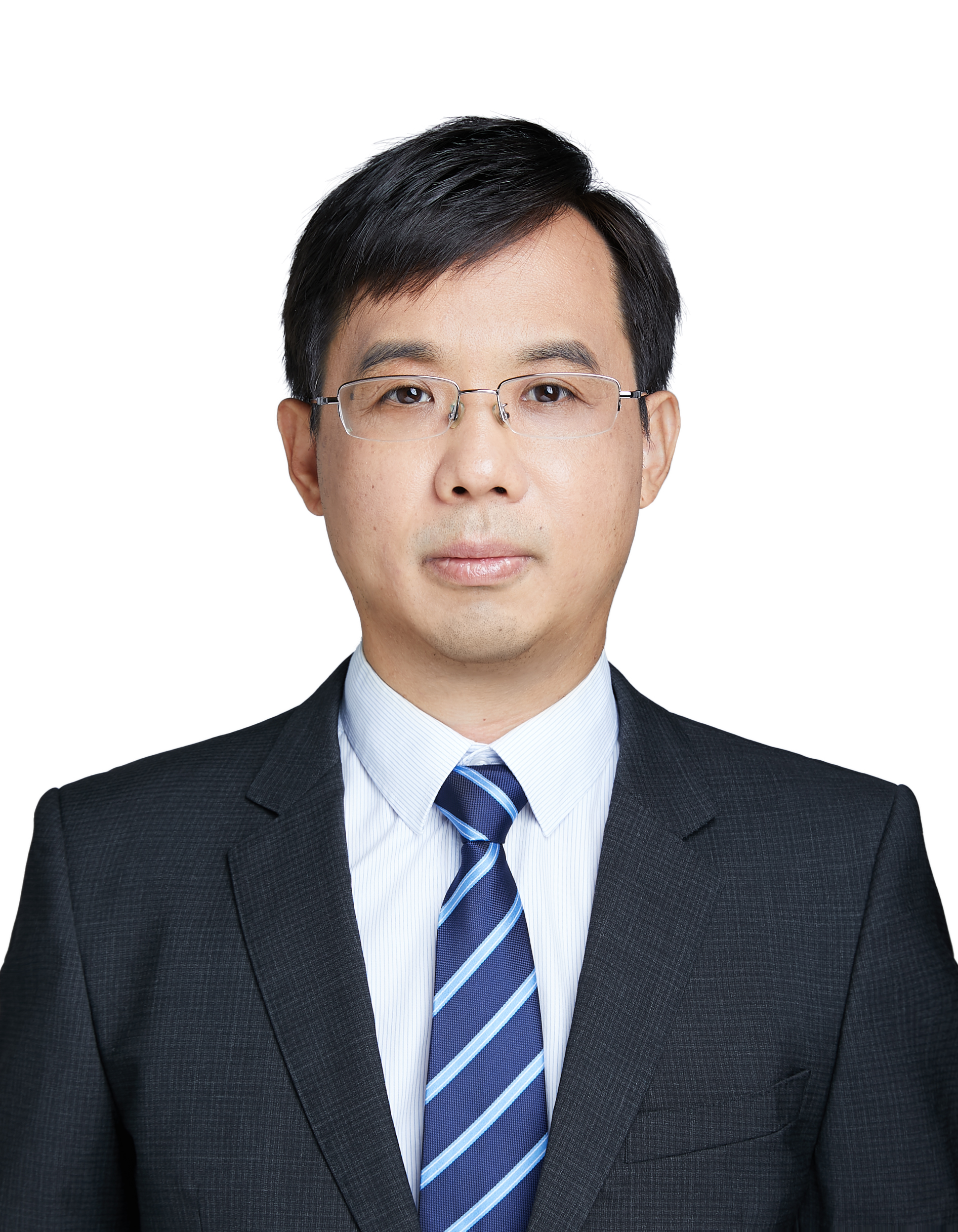}}]
{Zesong Fei} (Senior Member, IEEE) received the Ph.D. degree in electronic engineering from the Beijing Institute of Technology (BIT), Beijing, China, in 2004. He is currently a Full Professor with the Research Institute of Communication Technology, BIT.  Prof. Fei's research interests are in the area of wireless communications and signal processing. He has published more than 200 journal and conference papers. He serves as an Editorial Board Member for the IEEE Access.
\end{IEEEbiography}

\begin{IEEEbiography}[{\includegraphics[width=1in,height=1.25in,clip,keepaspectratio]{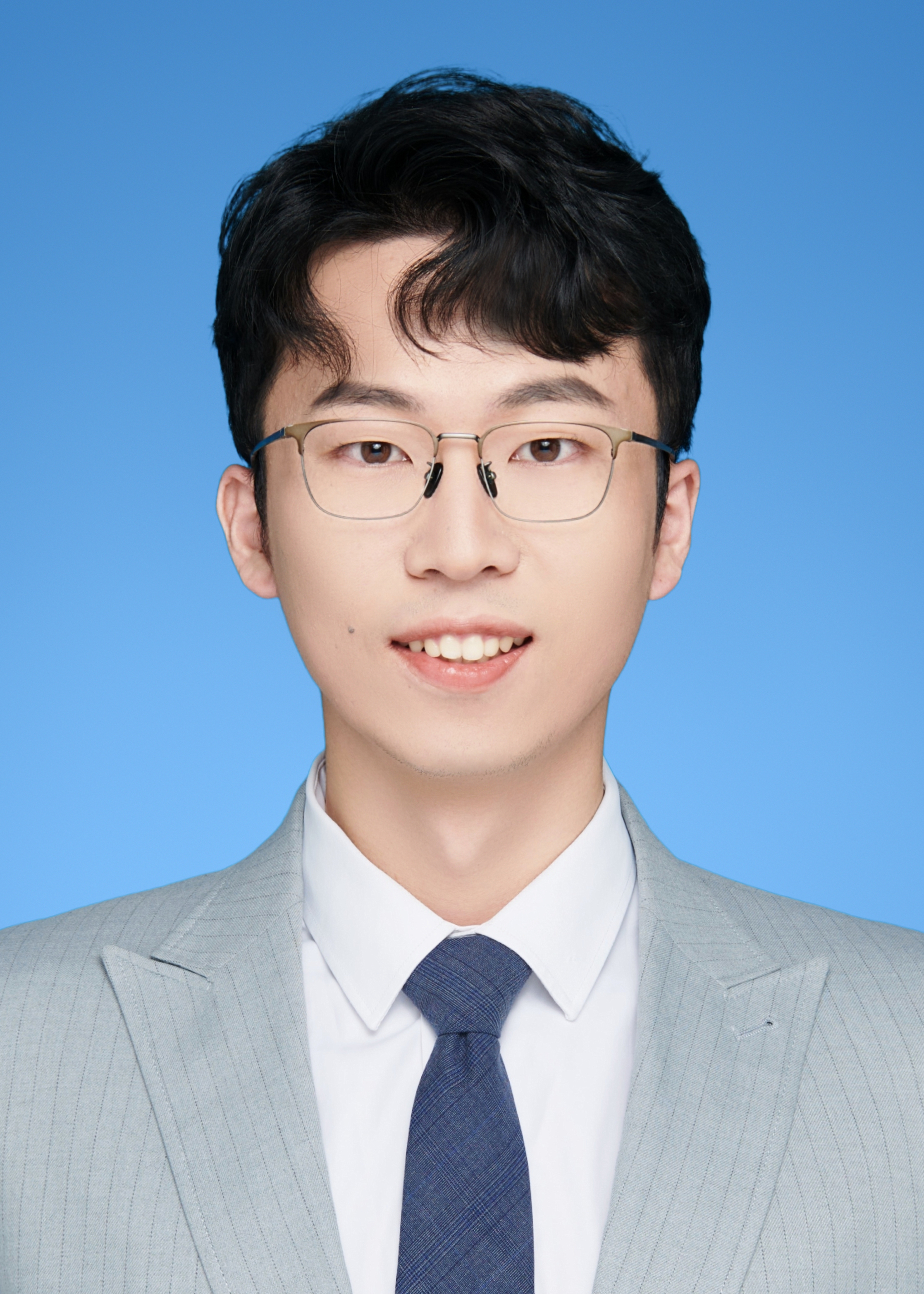}}]
{Xinyi Wang} (Student Member) received the B.S. degree in electronic engineering from Beijing Institute of Technology (BIT), Beijing, China, in 2017. He is currently working toward the Ph.D. degree with the Research Institute of Communication Technology, BIT. He was a receipt of the Best Paper Award in WOCC 2019. His research interests include integrated sensing and communications, UAV communications, intelligent reflecting surface, polar codes, and OTFS modulation.
\end{IEEEbiography}

\begin{IEEEbiography}[{\includegraphics[width=1in,height=1.25in,clip,keepaspectratio]{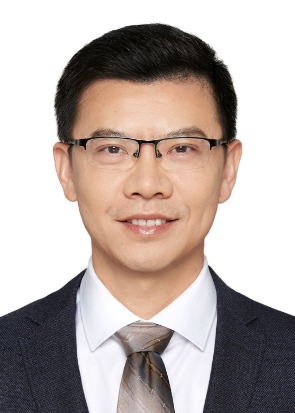}}]
{Nan Wu} (Member, IEEE) received the B.S., M.S., and Ph.D. degrees from the Beijing Institute of Technology (BIT), Beijing, China, in 2003, 2005, and 2011, respectively. From 2008 to 2009, he was a Visiting Ph.D. Student with the Department of Electrical Engineering, Pennsylvania State University, USA. He is currently a Professor with the School of Information and Electronics, BIT. His research interests include signal processing in wireless communication networks. He was a recipient of the National Excellent Doctoral Dissertation Award by MOE of China in 2013. He serves as an Editorial Board Member for the IEEE Wireless Communications Letters.
\end{IEEEbiography}

\begin{IEEEbiography}[{\includegraphics[width=1in,height=1.25in,clip,keepaspectratio]{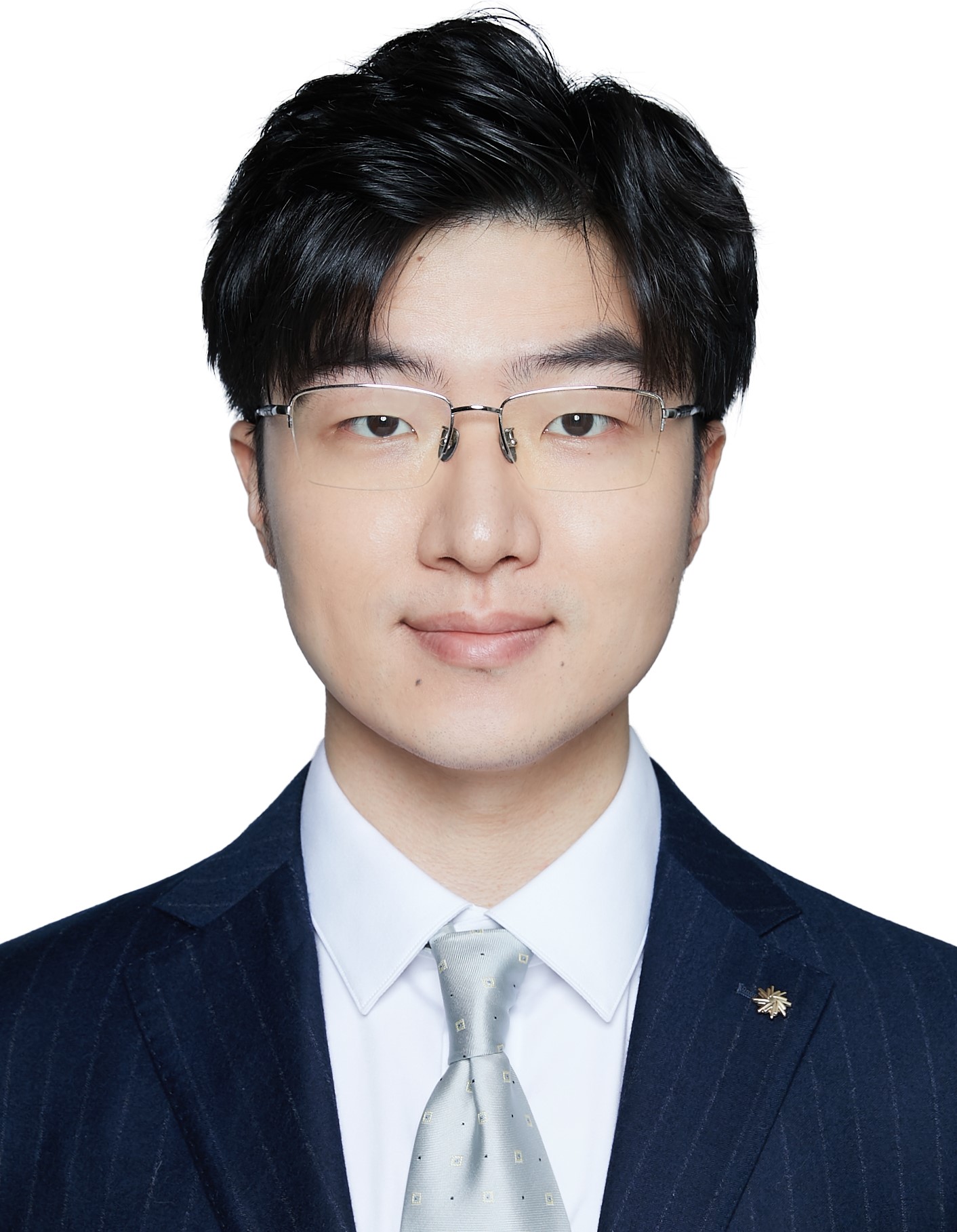}}]
{Jingxuan Huang} received the B.S. and Ph.D. degrees in electronics engineering from Beijing Institute of Technology (BIT), Beijing, China, in 2016 and 2021, respectively. He is currently a Post-Doctoral Fellow with the School of Information and Electronics, BIT. His research interests include channel coding and modulation, vehicular communications, and integrated sensing and communications.
\end{IEEEbiography}

\begin{IEEEbiography}[{\includegraphics[width=1in,height=1.25in,clip,keepaspectratio]{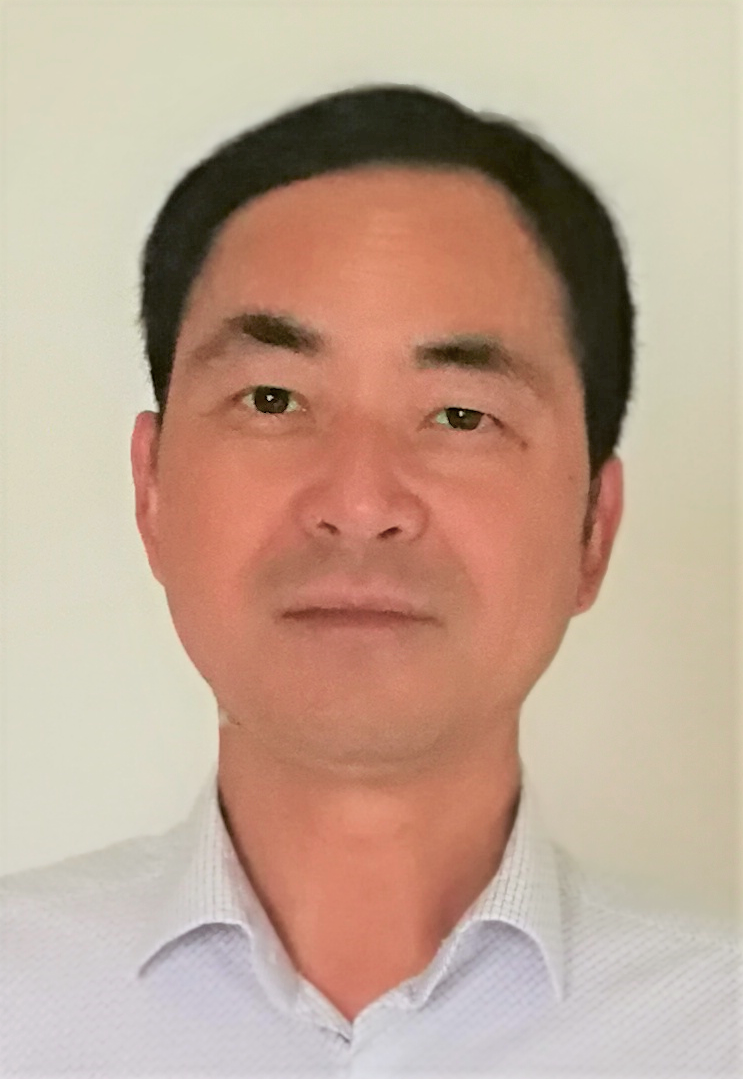}}]
{J. Andrew Zhang} (Senior Member, IEEE) received B.Sc. degree from Xi'an JiaoTong University, China, in 1996, M.Sc. degree from Nanjing University of Posts and Telecommunications, China, in 1999, and Ph.D. degree from the Australian National University, in 2004. Currently. He is an Associate Professor in the School of Electrical and Data Engineering, University of Technology Sydney, Australia. Dr. Zhang's research interests are in the area of signal processing for wireless communications and sensing. He has published 250+ papers in leading international Journals and conference proceedings, and has won 5 best paper awards for his work.
\end{IEEEbiography}

\vfill

\end{document}